\documentclass[namedreferences,hyperref,optionalrh,solaromanenum]{spr-sola}
\newcommand{\lya}{Ly$\alpha$\ }
\newcommand{\ha}{H$\alpha$\ }
\newcommand{\si}{Si~\textsc{iii}\ }

\makeatletter

\newcommand{\Rmnum}[1]{\expandafter\@slowromancap\romannumeral #1@}
\makeatother

\usepackage{txfonts}
\usepackage{hyperref}
\usepackage{threeparttable}

\usepackage[section]{placeins}
\usepackage{lmodern}
\usepackage{geometry}
\usepackage{natbib}
\usepackage{graphicx}                    
\usepackage{CJKutf8}
\usepackage{color}                       
\usepackage{multicol}
\usepackage{ulem}

\newcommand{\heii}{He~$\textsc{ii}$\ }

\chardef\us=`\_

\begin{document}
\begin{CJK}{UTF8}{gbsn}
\begin{sloppypar}
\begin{frontmatter}

\title{Understanding the Ly$\alpha$ Emission Observed by the Solar Disk Imager Aboard the Advanced Space-based Solar Observatory}


\author[addressref={1,2}]{\inits{Y.L.}\fnm{Yiliang}~\snm{Li}\orcid{0009-0006-1134-0205}}

\author[addressref={1}]{\inits{P.}\fnm{Ping}~\snm{Zhang}}

\author[addressref={1,2}]{\inits{Z.Y.}\fnm{Zhengyuan}~\snm{Tian}\orcid{0000-0002-2158-0249}}

\author[addressref={1,2},corref,email={lfeng@pmo.ac.cn}]{\inits{L.}\fnm{Li}~\snm{Feng}\orcid{0000-0003-4655-6939}}

\author[addressref={1,2}]{\inits{G.L.}\fnm{Guanglu}~\snm{Shi}\orcid{0000-0001-7397-455X}}

\author[addressref={1}]
{\inits{J.C.}\fnm{Jianchao}~\snm{Xue}\orcid{0000-0003-4829-9067}}

\author[addressref={1,2}]
{\inits{Y.}\fnm{Ying}~\snm{Li}\orcid{0000-0002-8258-4892}}

\author[addressref={1,2}]
{\inits{J.}\fnm{Jun}~\snm{Tian}\orcid{0000-0002-1068-4835}}

\author[addressref={3}]{\inits{K.F.}\fnm{Kaifan}~\snm{Ji}\orcid{0000-0001-8950-3875}}

\author[addressref={1}]{\inits{B.L.}\fnm{Beili}~\snm{Ying}\orcid{0000-0001-8402-9748}}

\author[addressref={1}]{\inits{L.L.}\fnm{Lei}~\snm{Lu}\orcid{0000-0002-3032-6066}}

\author[addressref={1,2}]{\inits{S.T.}\fnm{Shuting}~\snm{Li}\orcid{0000-0003-2694-2875}}

\author[addressref={1,2}]{\inits{J.S.}\fnm{Jiahui}~\snm{Shan}\orcid{0009-0001-4778-5162}}

\author[addressref={1,2}]{\inits{H.}\fnm{Hui}~\snm{Li}\orcid{0000-0003-1078-3021}}

\author[addressref={1,4}]{\inits{W.Q.}\fnm{Weiqun}~\snm{Gan}\orcid{0000-0001-9979-4178}}


\runningtitle{Understanding the \lya emission based on SDI observation}

\address[id={1}]{Key Laboratory of Dark Matter and Space Astronomy, Purple Mountain Observatory, Chinese Academy of Sciences, Nanjing 210023, China}

\address[id={2}]{School of Astronomy and Space Science, University of Science and Technology of China, Hefei 230026, China}

\address[id={3}]{Yunnan Observatories, Chinese Academy of Sciences, Kunming 650216, China}

\address[id={4}]{University of Chinese Academy of Sciences, Nanjing 211135, China}

\begin{abstract}

\vspace{0.03\textwidth} 
\indent The H~\textsc{i} Lyman-alpha (Ly$\alpha$) emission, with a wavelength of 1216~\AA, is the brightest solar ultraviolet (UV) line. However, comprehensive observations of the \lya emission line across the full solar disk remain limited. As part of the ASO-S mission, the Solar Disk Imager (SDI) has successfully captured full-disk images in the \lya band. Gaussian fitting of SDI's spectral response function (SRF) yields a full width at half maximum (FWHM) of approximately 85~\AA, which is significantly broader than the distance of \si line at 1206~\AA\ and the \lya line.
Thus, the emission contribution of \si to the SDI Ly$\alpha$ passband needs to be considered. For flares, in practice, we calculated the integrated intensity ratio $I$(Si~\textsc{iii})/$I$(Ly$\alpha$) by analyzing spectral observations from the SOLSTICE instrument. It yields values between 1.7\% and 14.6\%. Empirically, the ratio is proportional to the SXR flux.
Further analysis of spectral data from the SUMER instrument reveals that the ratio $I$(Si~\textsc{iii})/$I$(Ly$\alpha$) is approximately 0.5\% for prominences, 0.7\%--0.9\% for the inner disk, and 1.4\%--1.9\% close to the limb. These findings suggest that $I$(Si~\textsc{iii})/$I$(Ly$\alpha$)
is minimal for prominences and the inner disk, and the varying ratios across regions align with the center-to-limb variation of the Si~\textsc{iii} and \lya lines. Additionally, we compared \lya image intensity with 304~\AA, 1600~\AA, and 1700~\AA\ observations from AIA, as well as \ha from CHASE, in multiple regions (a prominence region, two active regions, and a quiet region). A relatively higher correlation of about 85\% is found between \lya and 304~\AA\ in active regions, whereas in the quiet region and prominence, their correlation coefficients are about 55\%.

\end{abstract}

%
\vspace{0.05\textwidth}
\keywords{Flares, Spectrum; Spectral Line, Intensity and Diagnostics; Prominences; Active Regions, Structure; Center-Limb Observations}

\end{frontmatter}

%
\section{Introduction}

Hydrogen is the most abundant element in the solar atmosphere \citep{1986A&A...154..154G} and throughout the solar-terrestrial environment. Consequently, the H~Lyman series, particularly the \lya emission line, play a critical role in studying the composition of the solar atmosphere and its radiative properties.
As the most prominent line in the vacuum ultraviolet~(VUV) range, the irradiance of the hydrogen \lya emission primarily arises from cool material in the solar system, where resonant scattering excites atomic hydrogen \citep{curdtLyMathsfAlpha2008}.
The \lya line serves as a significant source of hydrogen resonance excitation in both planetary and cometary atmospheres, as well as within the heliosphere \citep{ninaElectronProductionSolar2014}. \lya profiles have been extensively investigated, with high spectral-resolution observations achieved several decades ago \citep{Lemaire1978ApJ...223L..55L}. 
The profile typically exhibits a characteristic bimodal structure, where the line center appears as a self-absorbing feature due to resonant scattering effect. Comparatively recent findings indicate that the two peaks in the \lya profile are asymmetric, with the blue peak exhibiting a greater intensity than the red peak \citep{lemaireHydrogenLyLy2015a}. 
Observations of the \lya profile have consistently faced challenges due to strong, narrow geocoronal absorption at the line center. While the Solar Ultraviolet Measurements of Emitted Radiation (SUMER) aboard the Solar and Heliospheric Observatory (SOHO) addressed this issue by observing from the first Lagrange point \citep{lemaireFirstResultsSUMER1997}, gaps remain in imaging within the \lya passband. Furthermore, regular full-disk imaging observations were virtually nonexistent.

With the successful launch of the Advanced Space-based Solar Observatory (ASO-S: \citealp{gan2019RAA....19..156G}; \citealp{ganChineseSolarObservatory2022}) on 9 October 2022, full-disk imaging in the \lya band has become achievable. The scientific objective of ASO-S, the first comprehensive space-based solar observatory in China, is to investigate the formation, evolution, interaction, and interrelationships of the solar magnetic field, coronal mass ejections (CMEs), and solar flares \citep{ganAdvancedSpaceBasedSolar2023}. The Lyman-alpha Solar Telescope (LST: \citealp{LiFeng2019RAA....19..162F}; \citealp{2019HuiLiRAA....19..158L}; \citealp{chenInflightPerformanceCalibrations2024}) is one of the payloads aboard ASO-S, providing uninterrupted observations in the \lya passband from the disk center out to 2.5 solar radii. The Solar Disk Imager (SDI: \citealp{2019HuiLiRAA....19..158L}) is a key instrument of the LST, operating in the \lya passband with an aperture of 68~mm, a field of view (FOV) of $38.4^\prime$, and a spatial resolution of approximately $9.5^{\prime \prime}$ \citep{chenInflightPerformanceCalibrations2024}.

As the center of the \lya line forms in the lower transition region and its wings in the chromosphere \citep{1981ApJS...45..635V}, its radiation plays a crucial role in revealing different solar features. Previous studies have primarily concentrated on the irradiance and spectral characteristics of \lya line. \cite{1971SoPh...21..392G} was the first to observe \lya emission in the corona and upper chromosphere. Subsequent full-disk \lya profiles were obtained using data from Skylab \citep{1976JGR....81.3465N} and the Orbiting Solar Observatory (OSO~8: \citealp{1996SoPh..168...37B}). High-resolution observations of \lya spectra in specific solar regions were achieved through the High-Resolution Telescope and Spectrograph (HRTS: \citealp{1979ApJ...230..924B}) during rocket flights and via the Ultraviolet Spectromagnetograph (UVSP) aboard the Solar Maximum Mission (SMM: \citealp{1988ApJ...329..464F}). Additionally, the SUMER instrument enabled targeted observations in the \lya passband through raster scanning \citep{lemaireFirstResultsSUMER1997}. The Solar Radiation and Climate Experiment (SORCE: \citealp{Rottman2005}) has been used in obtaining spectral irradiance data across ultraviolet (UV) windows, including the \lya line, through the SOLar-Stellar Irradiance Comparison Experiment~\textsc{ii} (SOLSTICE~\textsc{ii}: \citealp{mcclintockSolarStellarIrradiance2005}). Recently, high-resolution \lya spectral observations have been achieved with the Chromospheric Lyman-Alpha Spectro-Polarimeter (CLASP: \citealp{2017ApJ...839L..10K}). For imaging purposes, the Very high Angular resolution ULtraviolet Telescope (VAULT: \citealp{Vourlidas2010SoPh..261...53V}) has revealed the \lya atmosphere, while the Extreme Ultraviolet Imager (EUI: \citealp{2020RochusA&A...642A...8R}) on the Solar Orbiter (SolO: \citealp{2020MullerA&A...642A...1M}) and ASO-S/SDI can be employed to capture images of specific regions and the entire solar disk, respectively.

When studying \lya profiles or irradiance, the \si emission line (1206~\AA) is generally taken into account \citep{curdtLyMathsfAlpha2008,tianHYDROGENLyaLyv2009,curdtSUMERLyLine2010}. 
The \si line typically forms in the upper transition region at a temperature of about 70,~000~K \citep{curdtSUMERLyLine2010}, which is slightly higher than the formation height of Ly$\alpha$. 
\cite{woods2004} has compared the variation of Si~\textsc{iii}, the \lya core and wing during the X17 flare in 2003. The \si emission showed the most pronounced rise by a factor of 17 during the impulsive phase, while wings and core of \lya increased by more than a factor of 2 and by 20\%, respectively. Yet research on \si irradiance relative to \lya in quiet sun or prominence regions has been less extensive. In this work, we investigated $I$(Si~\textsc{iii})/$I$(Ly$\alpha$) using SOLSTICE data and SUMER observations and presented the center-to-limb variation of \lya as observed by LST/SDI. Furthermore, we emphasized that multi-wavelength studies are also significantly meaningful for understanding the characteristics of the solar atmosphere and the interrelationships between different emission lines including Ly$\alpha$.

This paper is organized as follows: In Section~\ref{sec:SDI_SRF}, we present the spectral response function (SRF) of SDI and the full width at half maximum (FWHM) obtained through fitting. Section~\ref{sec:SDI and EUI} compares the \lya images captured by LST/SDI and SolO/EUI. In Section~\ref{sec:ratio of Si III to Lya}, we discuss $I$(Si~\textsc{iii})/$I$(Ly$\alpha$) in detail for flares in Section~\ref{sec:ratio in flare} and for quiet sun and prominences in Section~\ref{sec:ratio in prominence}. 
The differences in multi-bandpass images and the corresponding image intensity correlations are shown in Section~\ref{sec:multiple bands}.
Finally, we provide the summary and brief discussion of our findings in Section~\ref{sec:discussions and conclusions}.

\section{SDI Full-disk Imaging Observation and Spectral Response Function}
\label{sec:SDI_SRF}

\begin{figure}
    \centerline{ \includegraphics[height=0.4\textwidth,clip=]{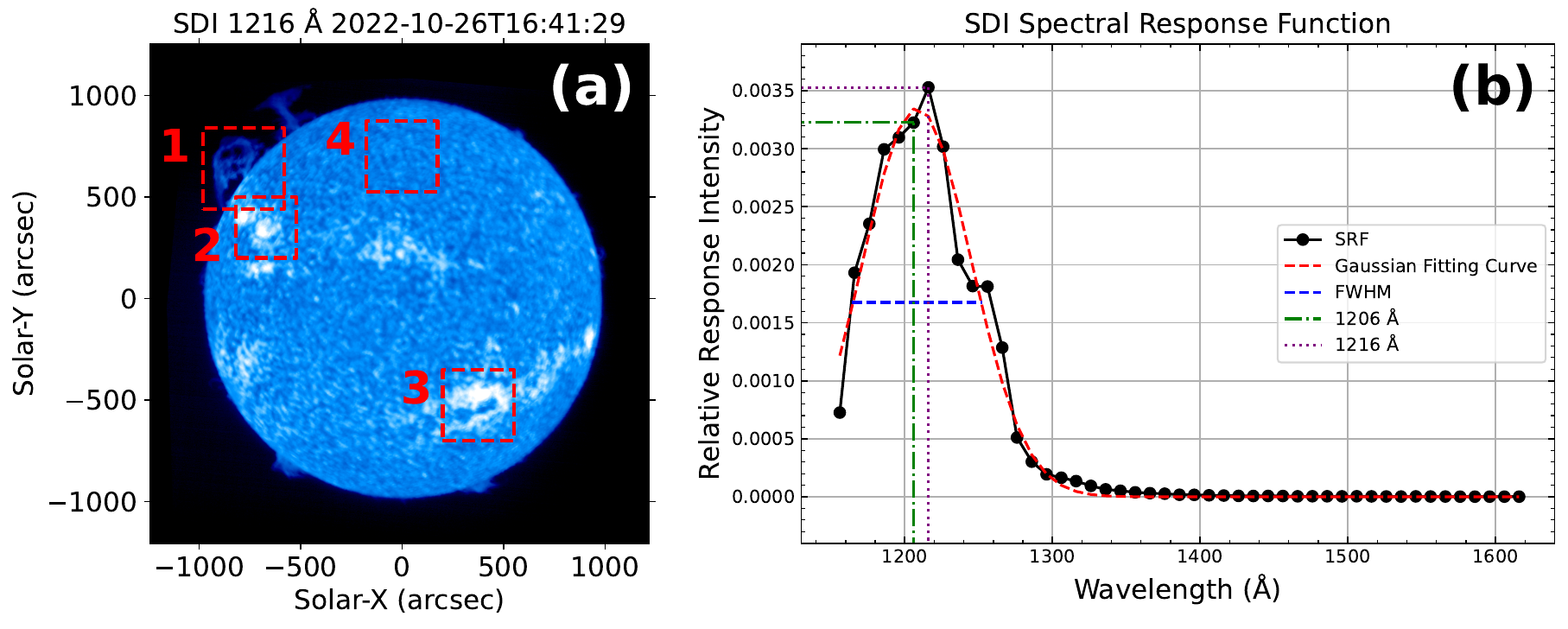}
    }
    \caption{\textbf{(a)} The full-disk \lya image observed by LST/SDI on 26 October 2022, at 16:41:29 UTC. The \textit{red dashed boxes} indicate the regions used for multi-band image comparison: Box~1: prominence region (PR); Box~2: active region with a filament (AR1); Box~3: active region with two filaments (AR2); Box~4: quiet region (QR). The specific information can be found in Table~\ref{tab:FOV of quiescent regions}. \textbf{(b)} The spectral response function of SDI. The \textit{black solid line} is about the curve the relative response intensity follows as wavelength changes. The Gaussian fitting curve is depicted by the \textit{red dashed line}, with the FWHM represented by the \textit{horizontal blue dashed line}. The data points at 1206~\AA\ and 1216~\AA\ are represented by \textit{green dash-dotted} and \textit{purple dotted lines}, respectively. The ratio of their corresponding SRF values is 0.914.}
    \label{fig:SDI_SRF}
\end{figure}

The Solar Disk Imager (SDI) is the first scientific instrument that achieved regular full-disk imaging in the Lyman-alpha passband \citep{chenInflightPerformanceCalibrations2024}. Shortly after the successful launch of ASO-S, SDI delivered its first set of full-disk images on 26 October 2022. 
Panel~a of Figure~\ref{fig:SDI_SRF} presents an image captured by the SDI on the day of its first opening, displaying structures such as prominences and active regions, as indicated by the red dashed boxed area. Further details will be discussed in Section~\ref{sec:multiple bands}.

The spectral response function (SRF) reflects the system transmission including optics and detectors.
The SRF of the SDI is presented in panel~b of Figure~\ref{fig:SDI_SRF}.  
The black curve represents the relative response intensity as a function of wavelength (see also in
\citealp{chenInflightPerformanceCalibrations2024}), measured prior to launch using the SDI flight model (FM) with a 10~\AA\ sampling interval.
The peak response is observed at 1216~\AA\ (characteristic \lya line wavelength), indicated by the purple dotted line, while the secondary maximum occurred at 1206~\AA\ (characteristic \si line wavelength), marked by the green dash-dotted line. Consequently, the calculated SRF ratio of \si to \lya is approximately 0.914. Additionally, we conducted a Gaussian fitting, indicated by the red dashed curve. Our analysis yields a FWHM of approximately 85~\AA, as indicated by the length of the dashed blue line. This substantial FWHM suggests the presence of other emission lines contributing to the SDI passband. Further details regarding this will be discussed in Section~\ref{sec:ratio of Si III to Lya}.

\section{Comparison of the Ly\texorpdfstring{$\alpha$} \ \ Emission Observed by SDI and EUI}
\label{sec:SDI and EUI}

\begin{figure}
    \centerline{ \includegraphics[width=0.9\textwidth,clip=]{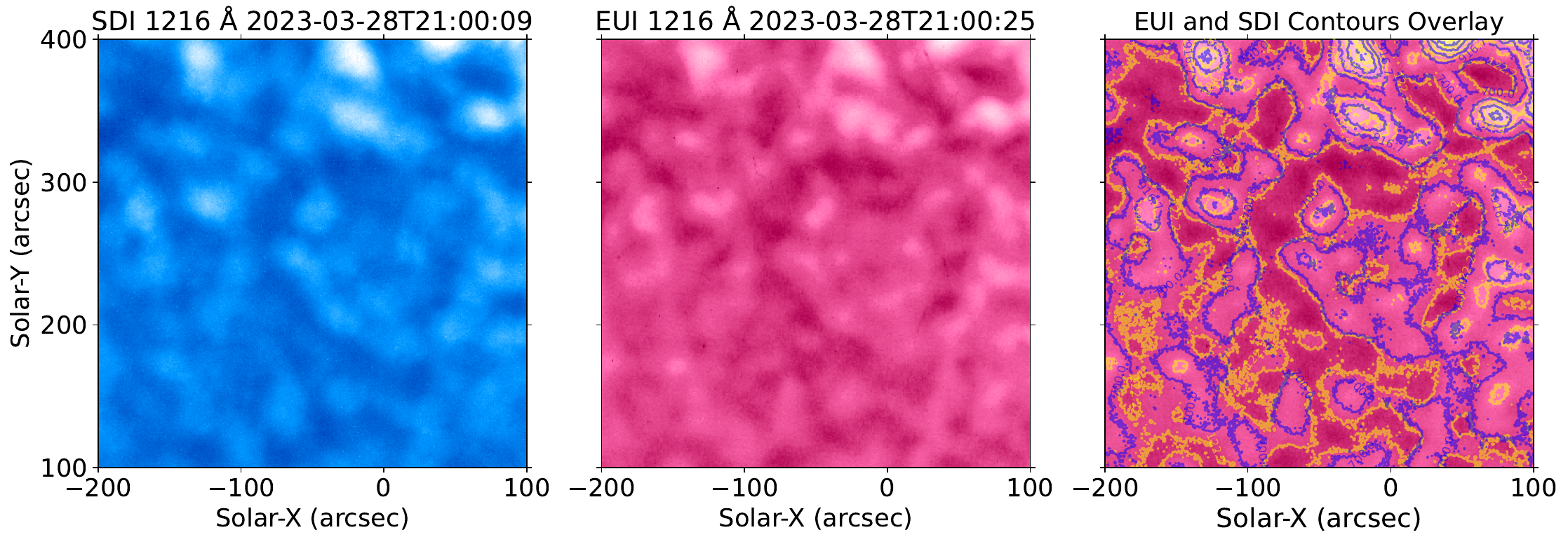}
    }
    \caption{Images of a relatively quiet region on the solar disk observed in the \lya passband on 28 March 2023, at 21:00 UTC. From left to right: the image captured by the SDI instrument, the image captured by the EUI instrument, and the contour overlay of both observations, with the \textit{blue contour} representing SDI and the \textit{yellow} one representing EUI.}    
    \label{fig-rela_SDISOLO}
\end{figure}

\begin{figure}
    \centering
    \includegraphics[width=0.9\textwidth]{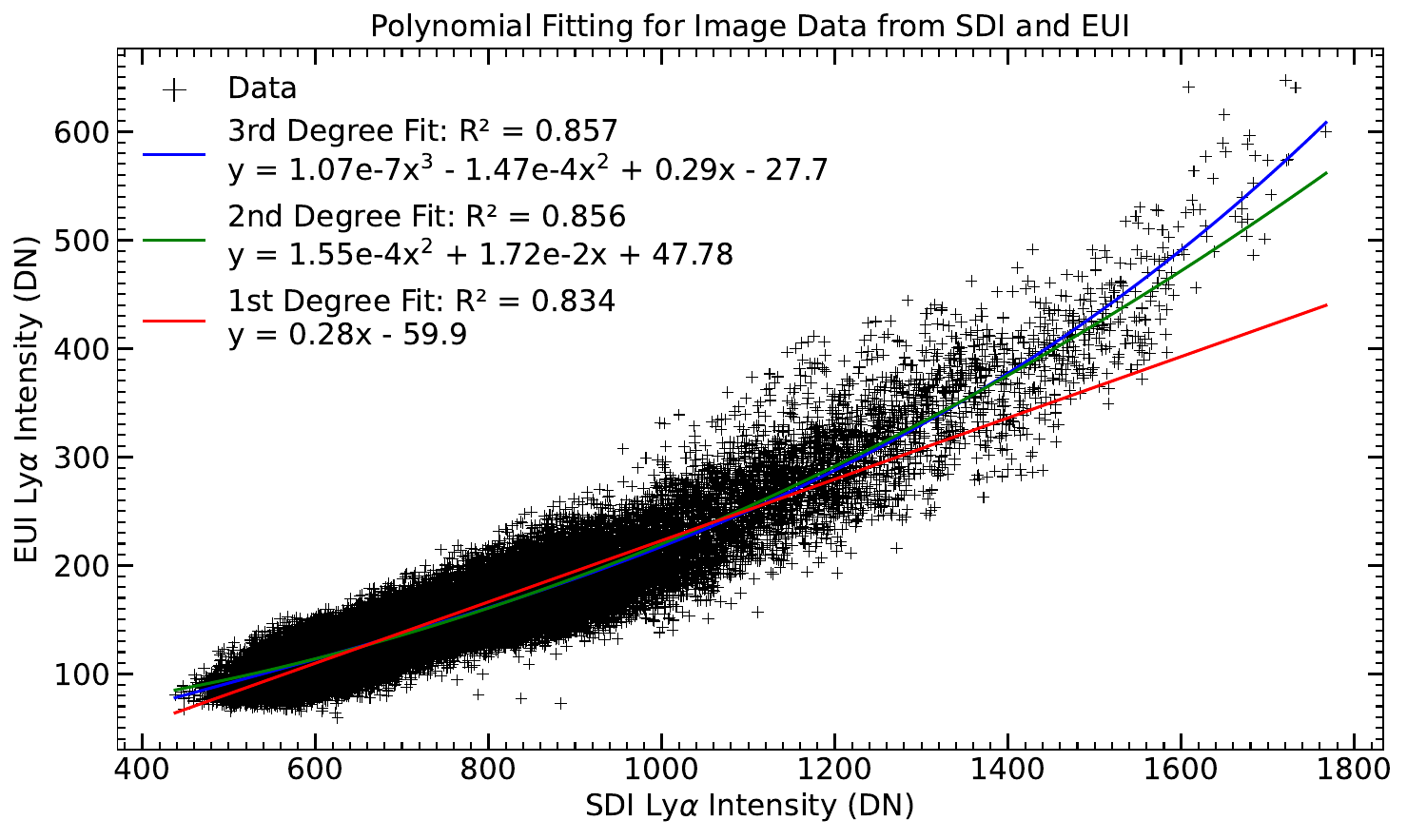}
    \caption{The scatter plot about the image intensity from the two instruments: SDI and EUI. The horizontal and normal axis is the count of the Ly$\alpha$ emission observed by the SDI and EUI, respectively. The \textit{red}, \textit{green}, and \textit{blue solid lines} separately represent the linear, quadratic and cubic fitting curves about the data points.}
    \label{fig:fit-SDISOLO}
\end{figure}

The Extreme Ultraviolet Imager (EUI: \citealp{2020RochusA&A...642A...8R}) is an instrument aboard the Solar Orbiter (SolO: \citealp{2020MullerA&A...642A...1M}), launched on 10 February 2020. One of EUI's two High Resolution Imagers is capable of capturing images in the \lya passband with a FOV of $1000^{\prime \prime} \times 1000^{\prime \prime}$. Both EUI and SDI are designed for imaging in the \lya passband, enabling a comparison of the solar \lya emission observed by the two instruments.

Due to significant difference in FOVs and slight difference in viewpoints, the raw images from SDI and EUI differ significantly, making alignment necessary for accurate comparison. For this analysis, we selected a relatively quiet region of the solar disk observed on 28 March 2023, when the Sun, SolO, and ASO-S were aligned almost along the same line. The two panels on the left of Figure~\ref{fig-rela_SDISOLO} display the aligned images of this region. The Spearman correlation coefficient between the two images is approximately 0.86, revealing a high degree of structural similarity. The right panel of Figure~\ref{fig-rela_SDISOLO} shows the contour overlay of the SDI and EUI images, demonstrating strong agreement, as the contours largely coincide.

We also conducted a statistical comparison of the image intensity characteristics by extracting the intensity from the SDI and EUI observations, presented in the scatter plot in Figure~\ref{fig:fit-SDISOLO}. The horizontal axis represents the intensity in the \lya passband, in digital numbers~(DN), observed by SDI, while the vertical axis shows the corresponding values observed by EUI. To further analyze the relationship between the two datasets, we applied multiple polynomial fits to the data. Linear, quadratic, and cubic fitted curves are represented by red, green, and blue solid lines, respectively.

The results indicate that both the quadratic and cubic fits provide a better match to the data compared to the linear fit, as reflected by the higher $R^2$ values. However, the overall trend of the scatter distribution suggests that the intensity of the EUI and SDI images is mainly in a linear relationship, especially when the value of DN is lower than 1200 for SDI. The fitting equation, displayed at the upper left of Figure~\ref{fig:fit-SDISOLO}, also shows this result. No matter quadratic or cubic fits, their nonlinear terms are several orders of magnitude smaller than the linear ones. The EUI intensity increases more rapidly relative to SDI at higher intensity level from the weak non-linear terms, which is maybe due to the difference of the SRFs for the two instruments.

\section{$I$(Si \textbf{\textsc{\textmd{iii}}})$/I$(Ly$\alpha$) for Different Solar Features}
\label{sec:ratio of Si III to Lya}

The \si emission line at 1206~\AA\ frequently appears in the solar atmosphere, and its wavelength is close to the \lya emission at 1216~\AA. Within the instrument's H~\textsc{i}~\lya channel, the \si emission line is detected alongside the primary \lya feature, exhibiting a notable high flux level \citep{woods1995}. In Section~\ref{sec:SDI_SRF}, we mentioned that the FWHM of the SDI's SRF is too broad to be solely attributed to the \lya emission. Therefore, the \lya images are inevitably blended with \si emission when observing the solar disk by SDI. This raises the question of how much the \si emission contributes to the emission in the SDI \lya passband.

To explore this, we analyzed the composition of emission lines, especially the \si line and the \lya line within the SDI \lya passband, in multiple flares and solar prominence events, etc., using observational data from SOLSTICE and SUMER.
For practical calculations, the integrated line emissions of \si and \lya are computed, respectively. Subsequently, their ratio $I$(Si~\textsc{iii})$/I$(Ly$\alpha$) is given.

\subsection{$I$(Si \textbf{\textsc{\textmd{iii}}})$/I$(Ly$\alpha$) for Flares Derived from SOLSTICE Observations}
\label{sec:ratio in flare}

\begin{figure}
    \centerline{
    \includegraphics[width=1.05\textwidth]{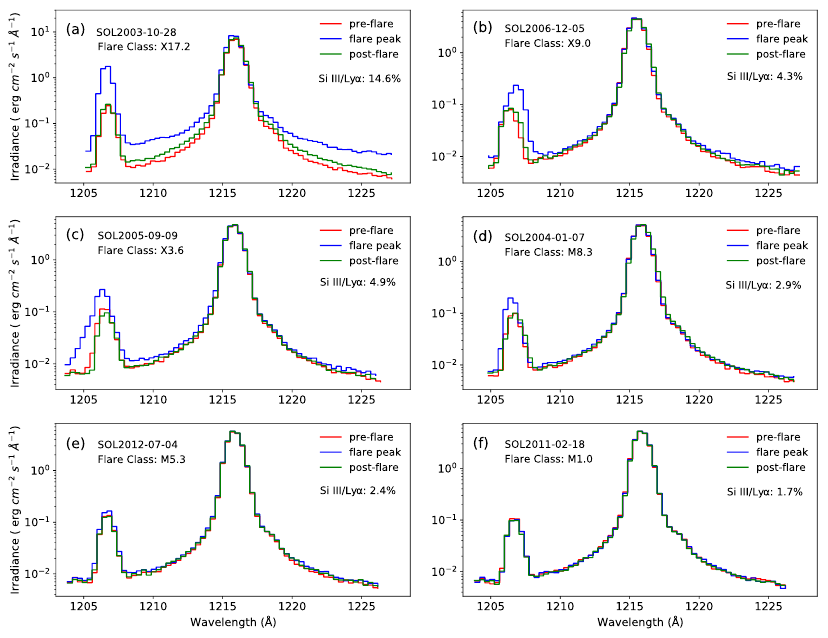}
    }
    \caption{Spectral profiles around the wavelength 1216~\AA\ at different M-level or X-level duration. In the \textit{upper left corner} of each panel, there are the dates of event occurrence and flare class. And in the right, Si~\textsc{iii}/Ly$\alpha$ represents the ratio of the integrated intensity from \si and \lya emission line. The \textit{red}, \textit{blue}, and \textit{green lines} of each panel are the spectra during pre-flare, flare peak and post-flare period, respectively.}
    \label{fig:the ratio of M-level flares}
\end{figure}

\begin{figure}
    \centering
    \includegraphics[width=0.8\textwidth]{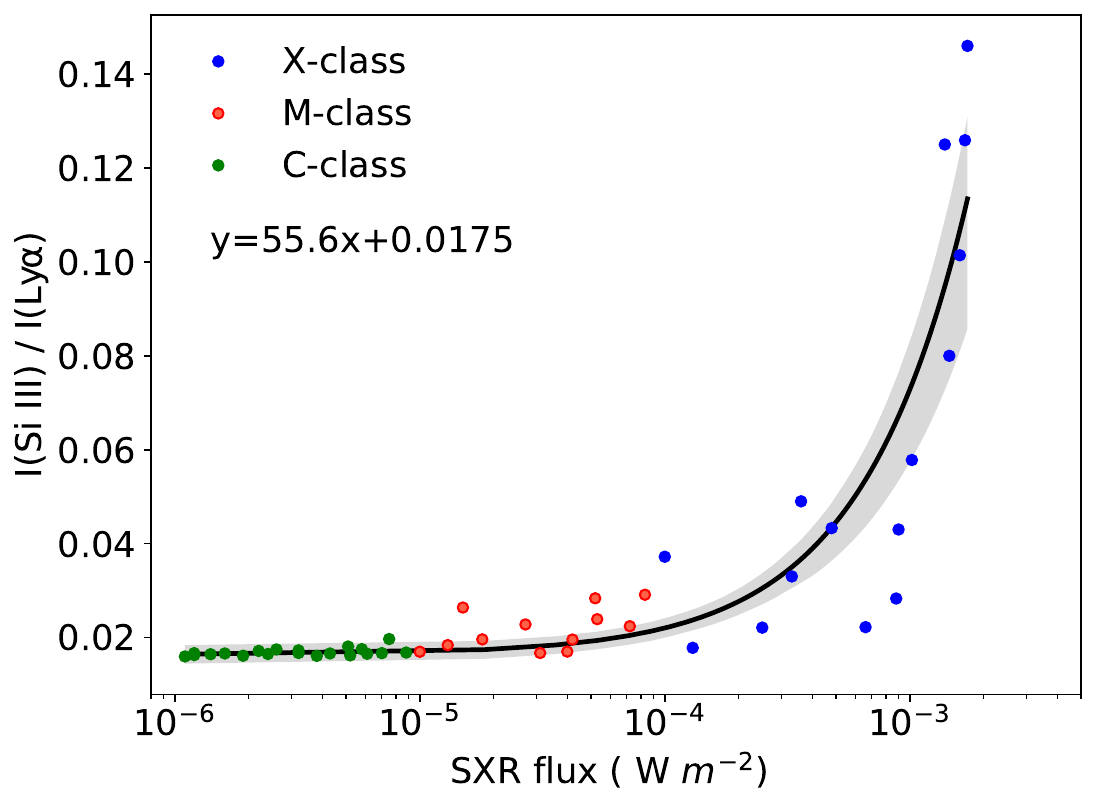}
    \caption{The scatter plot of the integrated intensity ratio $I$(Si~\textsc{iii})$/I$(Ly$\alpha$) is function of the flare SXR flux during 2003--2012. The \textit{solid circles} in \textit{blue}, \textit{red}, and \textit{green} represent X, M, and C-class flares, respectively. The samples are not only at peak time but also before and after the peak time. The \textit{black solid line} represents the linear regression curve, which is obtained by fitting the data points. The \textit{shaded area} represents the 95\% confidence interval of the linear regression.}
    \label{fig:ratio VS SXR flux}
\end{figure}

The Solar Radiation and Climate Experiment (SORCE: \citealp{Rottman2005}; \citealp{woodsOverviewSolarRadiation2021}), launched in January 2003, is a NASA-sponsored satellite mission designed to measure total solar irradiance (TSI) and solar spectral irradiance (SSI) across a wide range of wavelengths, including X-rays, ultraviolet (UV), visible, and infrared passbands. SORCE carries four primary instruments: the Spectral Irradiance Monitor (SIM), the SOLar-Stellar Irradiance Comparison Experiment~\textsc{ii} (SOLSTICE~\textsc{ii}), the Total Irradiance Monitor (TIM), and the XUV Photometer System (XPS).
The SOLSTICE~\textsc{ii} \citep{mcclintockSolarStellarIrradiance2005} instrument is a follow-up to the earlier SOLSTICE~\textsc{i} instrument \citep{rottmanSolarStellarIrradiance1993} and consists of two redundant grating spectrometers. These spectrometers measure both solar and stellar UV irradiance in two spectral ranges: 115--180~nm (far-ultraviolet, FUV channel) and 170--320~nm (middle-ultraviolet, MUV channel). In solar observation mode the entrance slit, with size of 0.1~mm~$\times$~0.1~mm provides a spectral resolution of 0.1~nm.
For our study, we used the Full Cadence Lyman-$\alpha$ data, covering the wavelength range of 1203--1227~\AA, which is obtained in spectral scanning mode at a resolution of 0.1~nm. This scanning process occurs one hour per day, taking about 67 seconds to scan a complete \lya profile, yielding full-disk solar irradiance measurements with a temporal resolution of approximately one minute.

To quantify the effect of \si relative to Ly$\alpha$, we analyzed the full-disk \lya profiles during six flares using SOLSTICE~\textsc{ii} data. The spectral profiles around the \lya center wavelength (1216~\AA) are shown in Figure~\ref{fig:the ratio of M-level flares}, which only includes profiles taken during M-class and X-class flares. Each panel represents the full-disk profile for a flare. The red, blue, and green lines correspond to the pre-flare, flare peak, and post-flare phases, respectively. The flare time and classification are indicated in the upper left of each panel, while the ratio of the integrated intensity over spectral ranges of \si to \lya is provided in the middle right. The integrated ranges are 1205--1208~\AA\ for \si and 1209--1227~\AA\ for Ly$\alpha$, respectively.
A prominent peak near 1206~\AA, corresponding to the \si emission, is visible next to the \lya peak in each panel. The \si peak and the \lya wing irradiance increase significantly during the flare peak phase as shown in panel~a and return almost to pre-flare levels in the post-flare phase. Notably, the \lya core irradiance remains relatively constant throughout the flare, consisting with the analysis of \cite{Greatorex2023ApJ...954..120G}: the \lya enhancements of three M-class flares from FISM2 were consistently 2.5\%.
The flare levels are correlated with Soft X-ray (SXR) flux: higher SXR flux corresponds to stronger flare enhancement. 
From panels a--f, the flare class decreases from X17.2 to M1.0, and a corresponding reduction in the difference between flare peak and pre-/post-flare irradiance, with the magnitude of the \si enhancement diminishing as the flare class decreases. This trend supports the correlation between \si irradiance and SXR flux. Consequently, the ratio of the integrated intensity $I$(Si~\textsc{iii})/$I$(Ly$\alpha$) also decreases with declining flare class, ranging from 14.6\% at X17.2 to 1.7\% at M1.0. Thus, in smaller flares, the \si contribution can be neglected, but for M-class flares and above, its contribution may have to be taken into account to obtain a more accurate \lya intensity from SDI data.

In order to further quantitatively characterize how much of the contribution of Si~\textsc{iii} is, we respectively computed the $I$(Si~\textsc{iii})$/I$(Ly$\alpha$) at the flares' SXR peak time, before and after the peak time observed by SOLSTICE from 2003 to 2012. The results are depicted in Figure~\ref{fig:ratio VS SXR flux}. The horizontal axis represents the SXR flux, while the vertical axis shows the corresponding intensity ratio. To differentiate between flare classes, we used blue, red, and green circles to represent X-class, M-class, and C-class flares, respectively. A linear regression curve, denoted by the black line, is fitted to the scatter points, with the shaded region indicating the 95\% confidence interval. The empirical function derived from the linear regression is provided in the upper left corner of Figure~\ref{fig:ratio VS SXR flux}.
The fit demonstrates a strong correlation, particularly for smaller flares. The results indicate an almost proportional relationship between the SXR flux and the \si to \lya intensity ratio, which is consistent with our earlier findings. This empirical function allows us to predict the intensity ratio for future flares if the SXR flux is known. 

More accurate \lya radiation measurements can be obtained from SDI observations during large solar flares, if the SRF is taken into account.
The common method applies the convolution with the SRF to spectral profiles to derive the integrated intensity $I$(Si~\textsc{iii}) and $I$(Ly$\alpha$), respectively, when computing the ratio $I$(Si~\textsc{iii})$/I$(Ly$\alpha$).
Due to the 10~\AA\ spectral resolution of the SRF in Figure~\ref{fig:SDI_SRF} matches the 10~\AA\ separation between the \si and \lya emission features, we adopted a simplified treatment wherein the SRF is held constant across each spectral window. Specifically, the instrument response is fixed at 1206~\AA\ for the \si window and at 1216~\AA\ for the \lya window.
To isolate the \lya intensity from the SDI measurements by removing the \si contribution, the derived intensity ratio $I$(Si~\textsc{iii})$/I$(Ly$\alpha$) is scaled by a factor of $\delta=0.914$, which is the ratio of SRF values between 1206~\AA\ and 1216~\AA.
Even for the X17.2-class flare shown in Figure~\ref{fig:the ratio of M-level flares}, the ratio $I$(Si~\textsc{iii})$/I$(Ly$\alpha$) only decreases from 14.6\% to 13.3\% when scaled by the factor $\delta$. For M-class flares, this ratio exhibits a reduction of less than 0.2\%, demonstrating negligible variation. Consequently, the original $I$(Si~\textsc{iii})$/I$(Ly$\alpha$) measurement establishes an upper limit for the \si contribution in SDI observations, with the resultant discrepancy remaining relatively minor.

\subsection{$I$(Si \textbf{\textsc{\textmd{iii}}})$/I$(Ly$\alpha$) for Quiet Sun and Prominence Derived from SUMER Observations}
\label{sec:ratio in prominence}

\begin{figure}
    \centerline{
    \includegraphics[width=1.05\textwidth]{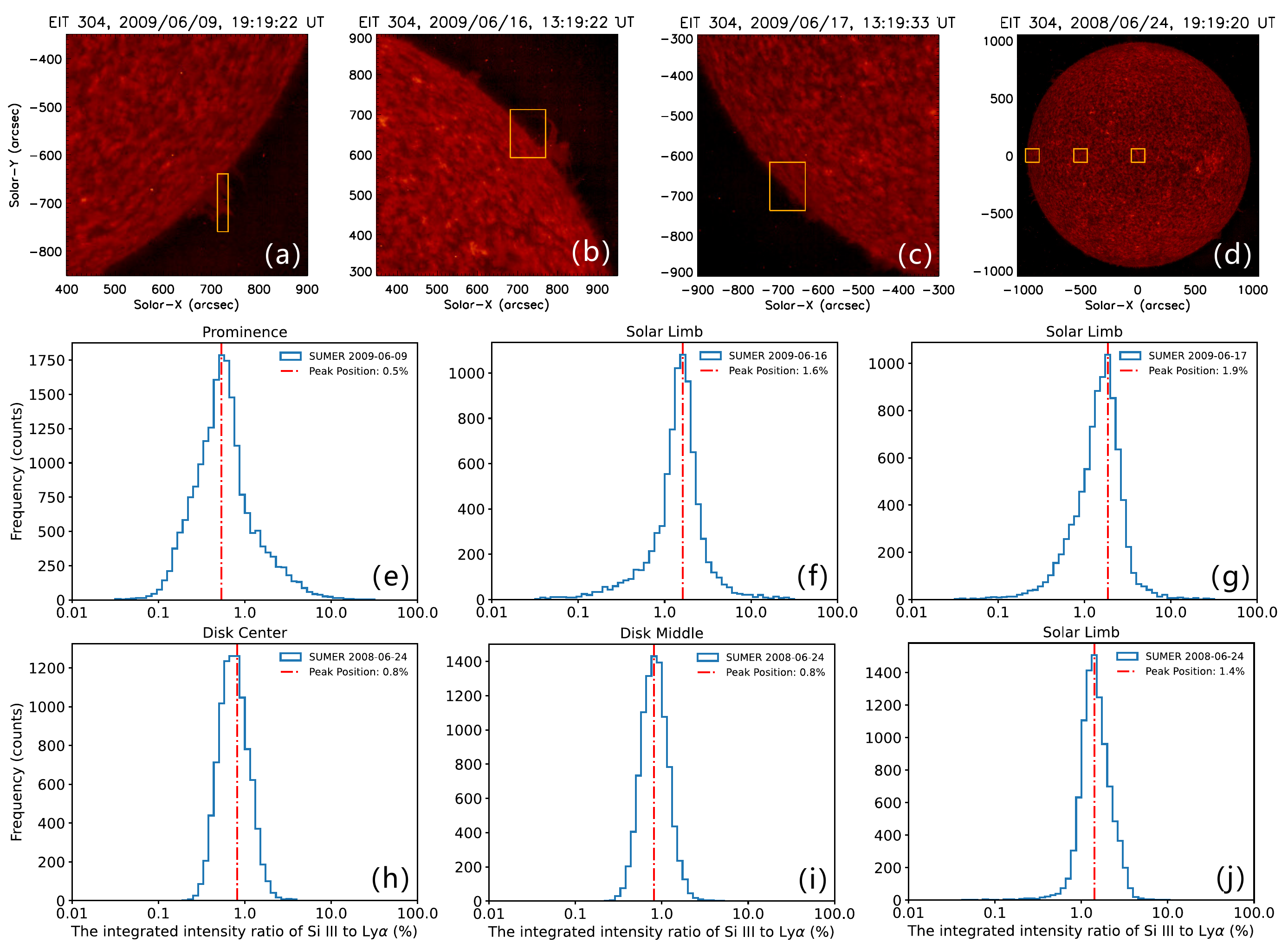}
    }
    \caption{\textbf{(a)}--\textbf{(d)}: The FOVs of SUMER scanning at disk outlined by \textit{orange rectangles} and \textit{squares}. The panel~a indicates the FOV of the prominence. The panel~b and c show the FOVs of the solar limb scanned by SUMER on 16 June and 17 June 2009. The \textit{squares} of panel~d separately indicate the FOVs of the disk center, disk middle, and solar limb from right to left. \textbf{(e)}--\textbf{(j)}: The distribution of the integrated intensity ratio $I$(Si~\textsc{iii})/$I$(Ly$\alpha$) in the multiple FOVs. The ordinate is the count in each bin, i.e. the frequency. The \textit{blue histogram} represents the distribution of the ratio and the \textit{red dash-dotted line} points out the peak position for every panel. The panels' title indicate the selected FOVs for performing the statistics. The observations by EIT and SUMER are not exactly simultaneous, and there is a time difference of the order of hours.}
    \label{fig:Ratio_SUMER}
\end{figure}

\begin{table}
\caption{The peak positions of the $I$(Si~\textsc{iii})/$I$(Ly$\alpha$) distribution on different selected regions in June 2008.}
\label{tab:Ratio_SUMER}
\begin{tabular}{c|c c c}\hline
    Date & 24 June 2008 & 25 June 2008 & 26 June 2008 \\\hline
    Disk Center & 0.8\% & 0.7\% & 0.7\% \\
    Disk Middle & 0.8\% & 0.8\% & 0.9\% \\
    Solar Limb & 1.4\% & 1.4\% & 1.9\% \\\hline
\end{tabular}
\end{table}

\begin{figure}
    \centerline{
    \includegraphics[width=1.0\textwidth]{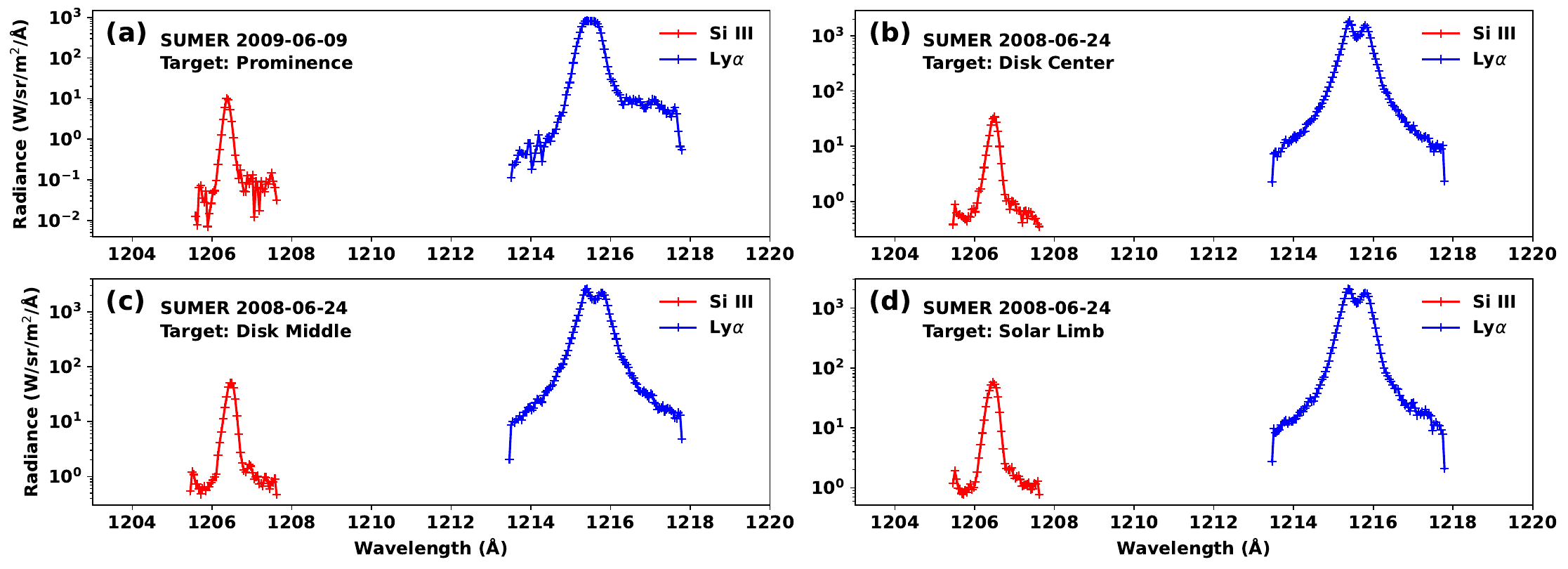}
    }
    \caption{The examples of integrated spectral profiles observed by SUMER. The targets in all panels are quiescent state. The \textit{red} and \textit{blue curve} indicate \si and \lya spectral windows in each panel, respectively.}
    \label{fig:SUMER_spectra_examples}
\end{figure}

\begin{figure}
    \centerline{
    \includegraphics[width=0.7\textwidth]{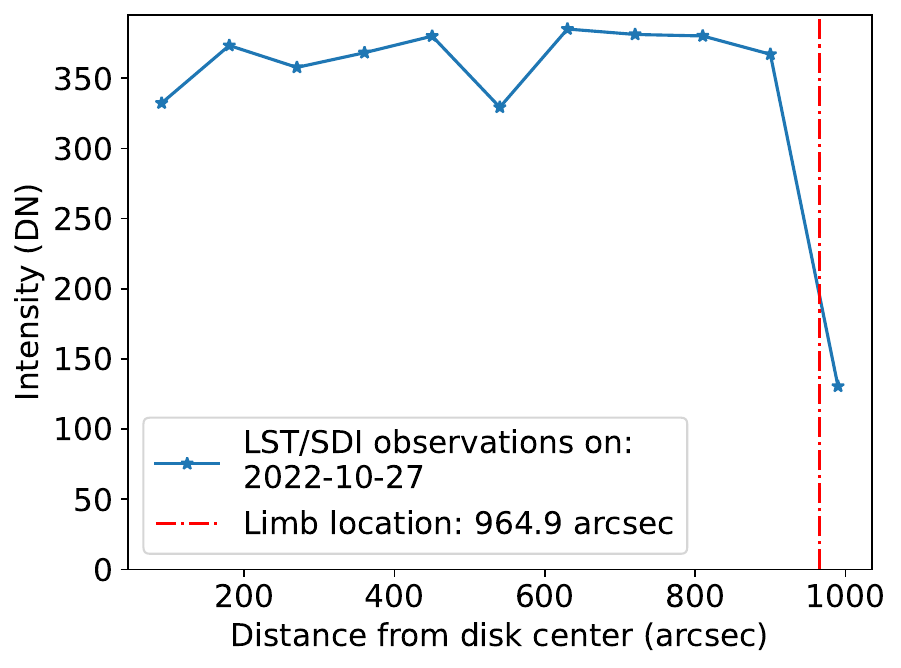}
    }
    \caption{The variation of \lya emission is function of distance from the disk center observed by SDI on 27 October 2022. The \textit{blue solid curve} indicates the variation of intensity. The \textit{red dash-dotted line} shows the location of solar limb.}
    \label{fig:limb_vary}
\end{figure}

The Solar Ultraviolet Measurements of Emitted Radiation (SUMER: \citealp{wilhelmSUMERSolarUltraviolet1995}) instrument aboard SOHO provided spectral observations in the wavelength range of 500~\AA\ to 1610~\AA\ since SOHO's launch in December 1995. SUMER's last observation was performed on 4 April 2017, and it has been put in hibernation since then.

On 9 June 2009, there was a quiet prominence observed using the raster scanning of SUMER with FOV of approximately $22^{\prime \prime}\times120^{\prime \prime}$ above the southwest limb \citep{curdtSUMERLyLine2010}, as shown in panel~a of Figure~\ref{fig:Ratio_SUMER}. The region outlined by the orange rectangle is the FOV of the SUMER scan. A $0.3^{\prime \prime}\times120^{\prime \prime}$ slit was used for each scan, with an exposure time of 14.5~seconds. The representative spectral profile acquired during the prominence observation by the SUMER instrument is presented in panel~a of Figure~\ref{fig:SUMER_spectra_examples}. 
The spectral profile exhibits two distinct observational windows corresponding to the \si and \lya channels, demonstrating morphological discontinuity when compared with the SOLSTICE continuum measurements in Figure~\ref{fig:the ratio of M-level flares}.
However, this discrepancy is expected to negligibly affect the calculation of $I$(Si~\textsc{iii})$/I$(Ly$\alpha$), as the omitted radiative contribution from the line wings is significantly smaller than that of the line core.
The radiance distributions of \lya and \si emissions observed in both the quiet Sun and a solar prominence, as reported by \cite{curdtSUMERLyLine2010}, motivate us to conduct a statistical analysis of the $I$(Si~\textsc{iii})/$I$(Ly$\alpha$) ratio. 
We employed the same observational data from the prominence study, which was obtained through photon flux reduction using a partially obscured aperture to prevent saturation of the \lya line (\citealp{curdtLyMathsfAlpha2008}; \citealp{curdtSUMERLyLine2010}).
The wavelength-integrated images of \si and \lya, not reproduced here but fully analyzed in \cite{curdtSUMERLyLine2010}, were derived from the prominence observation on 9 June 2009. The spectral windows span 1205.5–1207.6~\AA\ encompassing \si and 1213.5–1217.8~\AA\ centered on Ly$\alpha$, as shown in Figure~\ref{fig:SUMER_spectra_examples}.
Then we derived the distribution of $I$(Si~\textsc{iii})/$I$(Ly$\alpha$) in the prominence and generate a statistical histogram of the ratios, as shown in panel~e of Figure~\ref{fig:Ratio_SUMER}. The blue histogram represents the distribution of the ratios, while the red dashed line marks the location of the peak where the frequency reaches its maximum. Then the peak can represent the general ratio results. We noted that the peak position occurred at where the $I$(Si~\textsc{iii})/$I$(Ly$\alpha$) is 0.5\%. 
For comparison, we selected two additional observations in solar limb regions, with the FOVs of the SUMER observations as shown in panels~b and c of Figure~\ref{fig:Ratio_SUMER} and the resulting distributions displayed in panels~f and g.
Notably, the results in panel~f and g differ significantly from those in panel~e, with peak values of 1.6\% and 1.9\%, which are higher than the prominence result.
This suggests a substantial difference in radiance between \si and \lya when comparing the prominence with the limb, which is compatible with the findings of \cite{curdtSUMERLyLine2010}. 

Another SUMER observation, obtained in the \lya waveband on 24 June 2008 \citep{gunarQuietSunHydrogenLymanalpha2020}, and shown in panel~d of Figure~\ref{fig:Ratio_SUMER}, spark our research interest. 
Raster scans were performed across three quiet-Sun regions: the disk center, the midway between the disk center and the east limb, and the east limb itself, corresponding to the FOVs indicated in panel~d.
Each raster scan covered a FOV of $120^{\prime \prime}\times120^{\prime \prime}$, consisting of 80 slits with dimensions of $0.28^{\prime \prime}\times120^{\prime \prime}$, with a step size between slit positions of $1.5^{\prime \prime}$ and an exposure time of 15 seconds.
Panels b--d in Figure~\ref{fig:SUMER_spectra_examples} display the partial spectral profiles of this observation.
These SUMER data allow us to derive the $I$(Si~\textsc{iii})/$I$(Ly$\alpha$) distribution, as shown in panels~h, i, j of Figure~\ref{fig:Ratio_SUMER}, corresponding to the disk center, midway region (disk middle), and solar limb, respectively.
The peak positions are found to be 0.8\%, 0.8\%, and 1.4\%, respectively, indicating that the percentage effect of \si increases as the distance from the disk center grows.
Additionally, we also analyzed SUMER raster scans from 25 to 26 June 2008 (results not shown here), and the corresponding peak positions are summarized in Table~\ref{tab:Ratio_SUMER}. These three days' results are almost consistent: the farther from the disk center, the greater the integrated intensity ratio is.  

The results in the previous paragraph are reminiscent of the center-to-limb variation observed in earlier studies. 
\cite{tianHYDROGENLyaLyv2009} presented the center-to-limb variation of \si in their Figure~3, showing a gradual increase in \si radiance from the disk center to the limb, followed by a rapid decline above the solar limb (although the data samples cover only a small portion of the disk). 
In contrast, for the \lya emission line, the center-to-limb variation is typically described as remaining almost constant within disk, but with a comparably slower decrease than \si near and above the disk limb (\citealp{curdtLyMathsfAlpha2008}; \citealp{tianHYDROGENLyaLyv2009}). 
We also derived the center-to-limb variation of \lya using observations from LST/SDI for comparison. 
To avoid contamination from active regions, the sampling is performed at the solar equator. 
A series of squares, each measuring $90^{\prime \prime}\times90^{\prime \prime}$, are selected with a $90^{\prime \prime}$ step size, and the average intensity within each square is calculated, spanning from the disk center to beyond the west limb. 
The unprecedented center-to-limb variation of \lya across the entire disk is displayed in Figure~\ref{fig:limb_vary}, and it aligns well with previous conclusions---remaining nearly unchanged within the disk but rapidly decline near and above limb. 
Additional SDI observations from 28 to 31 October 2022 (not shown here) corroborate these findings.
Returning to the results shown in Figure~\ref{fig:Ratio_SUMER}, while the radiance of \si gradually increases from the disk center to the limb, the radiance of \lya remains practically invariant. 
Consequently, the $I$(Si~\textsc{iii})/$I$(Ly$\alpha$) slightly increases toward the limb, but above the disk limb, the ratio drops sharply as the \si radiance decreases more dramatically than that of Ly$\alpha$.
The ratio's behavior observed in prominence is consistent with the trend, further supporting the conclusion that \si radiance shows a pronounced decline above the limb, while \lya remains more stable.
On the other hand, the $I$(Si~\textsc{iii})/$I$(Ly$\alpha$) for the quiet Sun and prominence consistently remain relatively low, never exceeding 2\%. This indicates that the contribution of \si can be almost negligible.
Consistent with the procedure in Section~\ref{sec:ratio in flare}, the intensity ratio $I$(Si~\textsc{iii})/$I$(Ly$\alpha$) requires application of the scaling factor $\delta=0.914$ to compensate for the influence of SRF, enabling more precise determination of the \si emission contribution in SDI data. 
In the quiet Sun and prominence regions discussed previously, the \si contribution is expected to be further reduced, remaining constrained below 1.8\%. Consequently, the \si emission within the SDI passband and the SRF effect demonstrate negligible impacts.

\section{Comparison Between \texorpdfstring{\lya} \ Emission and Images at Other Wavelengths}
\label{sec:multiple bands}

\begin{table}
\caption{The FOVs of the four selected regions in Figure~\ref{fig:SDI_SRF}.}
\label{tab:FOV of quiescent regions}
\begin{tabular}{c|c|c|c|c}\hline
    Region & Box 1 (PR) & Box 2 (AR1) & Box 3 (AR2) & Box 4 (QR) \\\hline
    X (arcsec) & [-980, -580] & [-820, -520] & [200, 550] & [-175,175]\\
    Y (arcsec) & [440, 840] & [200, 500] & [-700, -350] & [525, 875]\\\hline
\end{tabular}
\end{table}

To examine the structures and surrounding environment of prominences and filaments, we selected various solar regions listed in Table~\ref{tab:FOV of quiescent regions} observed on 26 October 2022. Taking advantage of the full-disk \lya images provided by LST/SDI onboard ASO-S, we conducted a multi-wavelength study, comparing the imaging intensity of prominences, active regions with filaments, and quiet regions. The compared images are captured in Ly$\alpha$, H$\alpha$, and other multiple ultraviolet (UV) wavelengths (304~\AA, 1600~\AA, 1700~\AA).
These are analyzed in conjunction with \ha observations taken by the \ha Imaging Spectrograph (HIS: \citealp{qiuCalibrationProceduresCHASE2022}), part of the scientific payload aboard the Chinese \ha Solar Explorer (CHASE: \citealp{liChineseHaSolar2022}) satellite, as well as with observations by the Atmospheric Imaging Assembly (AIA: \citealp{lemenAtmosphericImagingAssembly2012}) aboard the Solar Dynamics Observatory (SDO), launched on 11 February 2010.

The SDI data underwent dark field subtraction and flat-field correction prior to spatial co-registration with AIA 304~\AA\ observations, with the method described in \cite{Wangyun2025ApJ...982..161W}. 
The cross-instrument aligned images described in this section maintain identical array dimensions, enabling direct pixel-by-pixel correlation analysis of their intensity distributions. The CHASE H$\alpha$ images are observed in raster scanning mode (RSM: \citealp{qiuCalibrationProceduresCHASE2022}) with a spatial resolution of 2$^{\prime\prime}$ at mode of $bin=2$ and pixel spectral resolution of approximately 0.024~\AA. CHASE data are corrected for dark-field, slit-image curvature, and flat-fielding before co-alignment with AIA 304~\AA\ images. 
To facilitate comparison with images from other wavelengths, we integrated the \ha spectral data along the wavelength direction. The selected integration range spans 0.4~\AA\ above and below the \ha line center (6562.8~\AA), based on the calibrated Level 1 RSM spectrum and spectral profiles (\citealp{Fang2022IntroductionTT}; \citealp{qiuCalibrationProceduresCHASE2022}).
AIA provides multiple 
high-resolution full-disk images with 1.5$^{\prime\prime}$ spatial resolution and 12~seconds temporal resolution. The resulting images are shown in Figure~\ref{fig:all-image}, with each column corresponding to the regions marked with squares in Figure~\ref{fig:SDI_SRF}: Box 1 (PR), Box 2 (AR1), Box 3 (AR2), and Box 4 (QR), while each row corresponding to various passbands mentioned previously. The details of these observations will be discussed in the following subsections.

\subsection{On Prominence}
\begin{table}
\caption{The correlation coefficients between each pair of passbands, as calculated from the PR images.}
\label{tab:correlation_prominence}
\begin{threeparttable}
\begin{tabular}{c|c c c}\hline
Waves & 304 VS \lya & H$\alpha$ VS \lya & H$\alpha$ VS 304 \\\hline
PR & 0.530 & 0.317 & 0.333 \\\hline
 Along Cut0 & 0.384 & 0.549 & 0.173 \\
\hline
\end{tabular}
\end{threeparttable}
\end{table}

\begin{figure}
    \centerline{
    \includegraphics[width=1.05\textwidth, clip=]{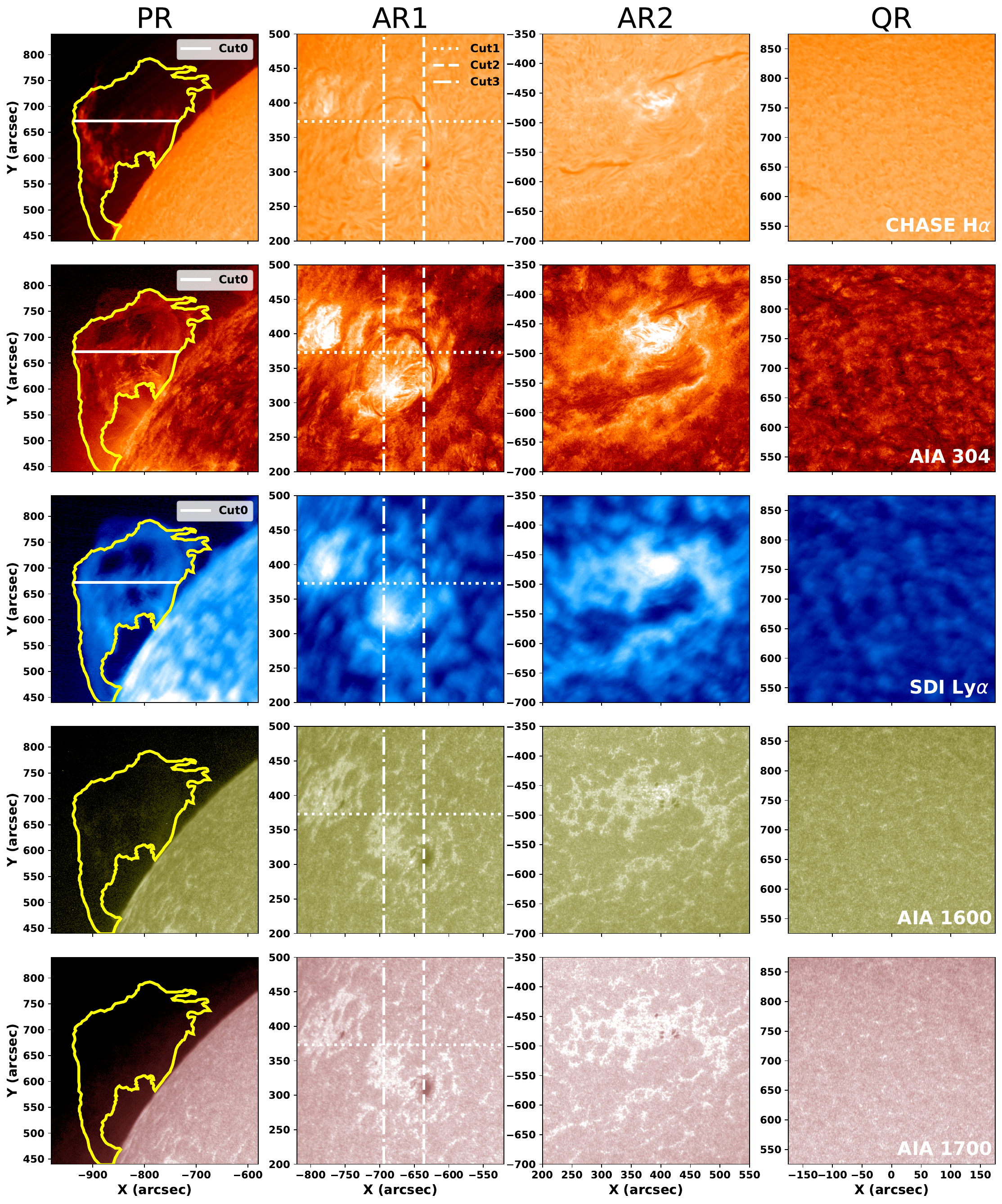}
    }
    \caption{Multi-wavelength images of the four selected regions (PR, AR1, AR2, QR). The four columns correspond to these regions from left to right, while the five rows represent different channels: CHASE H$\alpha$, AIA 304~\AA, SDI Ly$\alpha$, AIA 1600~\AA, and 1700~\AA\ from top to bottom. In the PR images, the main area where the eruptive prominence appears is outlined by \textit{yellow curves}, with a \textit{horizontal cut} marked by a \textit{white solid line} and labeled as Cut0 for further analysis. In the AR1 images, three cuts are similarly marked by \textit{white lines}: a \textit{horizontal cut} labeled Cut1, and \textit{two vertical cuts} labeled Cut2 and Cut3, from right to left.}
    \label{fig:all-image}
\end{figure}

\begin{figure}
    \centerline{
    \includegraphics[width=0.8\textwidth, clip=]{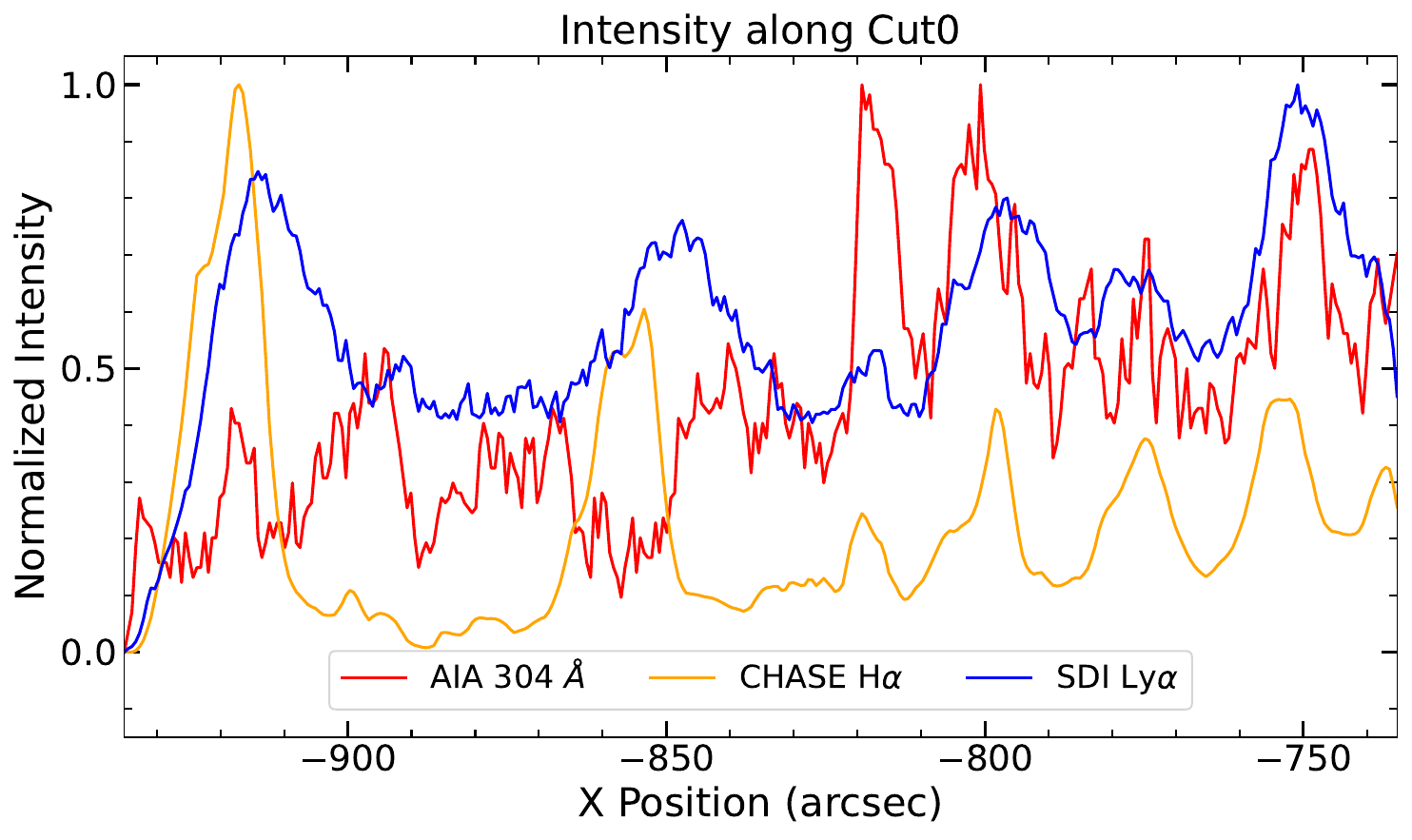}
    }
    \caption{The normalized intensity distribution along Cut0 in the PR images. The \textit{red}, \textit{orange}, and \textit{blue lines} represent the intensity at 304$\r{A}$, H$\alpha$ and \lya passbands, respectively.}
    \label{fig:cut_prominence}
\end{figure}

On 26 October 2022, a prominence eruption was observed at the northeastern limb. 
No associated CME was detected.
The eruption was captured by the LST/SDI at around 16:41 UT \citep{xueAssociationFailedProminence2024}, as shown in the left panel, middle row of Figure~\ref{fig:all-image}.
The SDI had a cadence of 10 seconds, but the data between 15:44 UT and 16:41 UT were lost. 
The upper two panels display the aligned images of the eruptive prominence, captured around 16:41 UT by CHASE in the \ha band and SDO/AIA in the \heii 304~\AA\ passband.
The main area of the eruptive prominence is outlined in yellow curves.
The prominence appears more extended in the \lya and 304~\AA\ passbands compared to the H$\alpha$, as the part of cooler prominence material is invisible in \ha for optically thinner but visible in \lya and \heii lines for optically thicker (\citealp{Heinzel2001ApJ...561L.223H}; \citealp{Schmieder2003A&A...401..361S}). Thus, more complete prominence structure is visible in both the 304~\AA\ and \lya passbands, with the 304~\AA\ band showing more detailed features.
In contrast, only part of the prominence structure is visible in H$\alpha$. As the prominence is invisible in the 1600~\AA\ and 1700~\AA, only three passbands (304~\AA, \lya and H$\alpha$) are considered at prominence region (PR) and the region outlined is used to calculate the correlation coefficients of image intensity between the different passbands.

The derived correlation coefficients are displayed in Table~\ref{tab:correlation_prominence}.
Since the calculated Pearson coefficients are largely similar to the Spearman coefficients, and considering the stability of the latter, only Spearman correlation values are included in this paper.
None of the correlations are statistically significant for PR, which is understandable given the visual differences between the prominence structures in the three images.
The highest similarity, and thus the maximum correlation coefficient of about 0.5, is found between the 304 \AA\ and \lya passbands, which is consistent with the overall structural resemblance of the prominence in the two passbands. 
The lowest correlation occurs between \ha and \lya with a value of 0.317, which is similar to the result between \ha and 304~\AA. 

More specifically, a horizontal cut, labeled Cut0, is selected across the prominence (marked by a white solid line) to extract intensity and calculate the correlation coefficients along the cut. 
The results, also displayed in Table~\ref{tab:correlation_prominence}, show a different pattern.
The highest correlation value is now found between the \ha and Ly$\alpha$, whereas the correlation between 304~\AA\ and \ha has significantly decreased.
Figure~\ref{fig:cut_prominence} illustrates the normalized intensity distribution along Cut0, with red, orange and blue lines representing 304~\AA, \ha and Ly$\alpha$, respectively.
Several inverse features are observed between \ha and 304~\AA, where one curve shows a peak while the other shows a valley. These inversions explain the lower correlation.
Conversely, the intensity distributions of \lya and \ha exhibit an overall consistency, with multiple peaks occurring simultaneously in both passbands, suggesting that several bright features of the prominence are captured along Cut0. While there are differences in smaller-scale structures, this overall similarity enhances the correlation between the two passbands.

\subsection{On Active Regions with Filaments}
\begin{table}
\caption{The correlation coefficients obtained from the AR1 images in Figure~\ref{fig:all-image}.}
\label{tab:correlation_filament}
\begin{tabular}{c|c c c c c}\hline
Waves & 304 VS \lya & \ha VS \lya & 304 VS \ha & 1600 VS \lya & 1700 VS \lya \\\hline
AR1 & 0.845 & 0.222 & 0.130 & 0.414 & 0.281 \\\hline
Cut1 & 0.767 & -0.065 & 0.135 & 0.533 & 0.348 \\
Cut2 & 0.867 & 0.334 & 0.406 & 0.172 & 0.216 \\
Cut3 & 0.885 & 0.720 & 0.639 & 0.690 & 0.640 \\\hline
\end{tabular}
\end{table}

\begin{figure}[!htb]
    \centerline{
    \includegraphics[width=1.02\textwidth,clip=,angle=0]{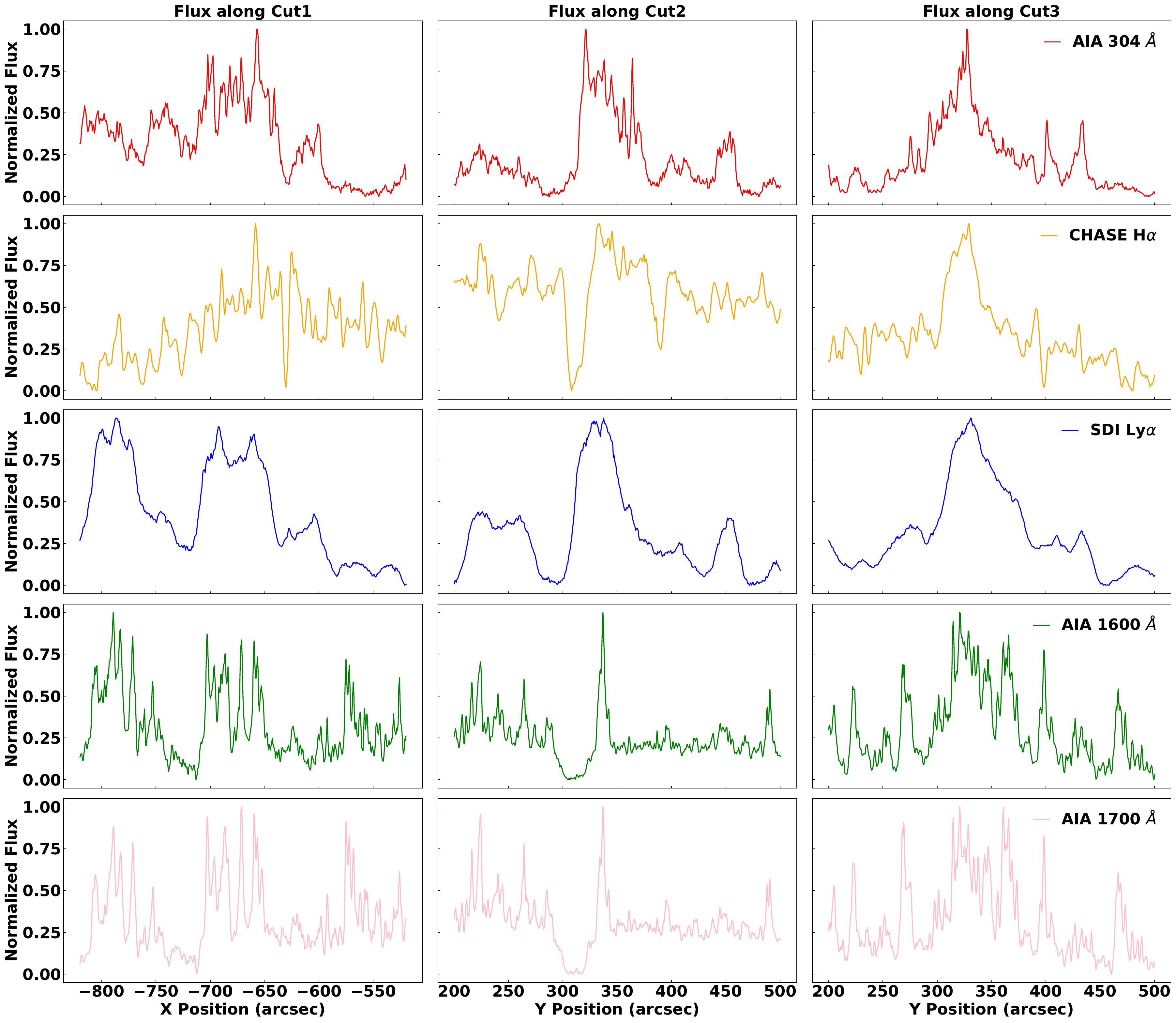}
    }
    \caption{The normalized intensity, along the cuts at AR1. The three columns, from left to right, are corresponding to the Cut1, Cut2 and Cut3. And in each column, the panels are respectively corresponding to the AIA 304~\AA, CHASE H$\alpha$, SDI Ly$\alpha$, AIA 1600~\AA, and 1700~\AA\ from top to bottom.}
    \label{fig:cut_filament}
\end{figure}

Around the time of the eruptive prominence, an active region containing a filament (AR1) is selected to compare the structural differences across multiple passbands. The second column of Figure~\ref{fig:all-image} shows the multi-wavelength observations of AR1. Due to the photons at 1600~\AA\ and 1700~\AA\ primarily originate from C~\textsc{iv} and the continuum (\citealp{lemenAtmosphericImagingAssembly2012}), these channels mainly display the features of upper photosphere and the region of temperature minimum, respectively, so the filament is not visible in either lines. However, a clear filament structure is observed in the other three wavelengths. The filament's absorbing feature appears wider in \lya and \heii 304 \AA\ than in H$\alpha$. This may be attributed to the fact that both \lya and \heii 304~\AA\ are resonant lines, both optically thicker than the H$\alpha$ line.

Analogous to the previous subsection, we calculated the correlation coefficients for the entire AR1, and the results are shown in Table~\ref{tab:correlation_filament}. The highest correlation is also between 304~\AA\ and Ly$\alpha$, with a value more than 0.8, which is highly significant. This suggests that the structures in the two passbands are closely resembled. 
The lowest correlation is found between 1700~\AA\ and Ly$\alpha$, likely because of the different heights of formation—while the Ly$\alpha$ emission normally originates from the chromosphere and lower transition region (TR), the 1700~\AA\ emission originates from the photosphere.

In AR1, we also selected three cuts through the filament, indicated by white lines in
Figure~\ref{fig:all-image}. The horizontal dotted line represents Cut1, while the two vertical dashed and dotted-dashed lines indicate Cut2 and Cut3, respectively.
The correlation coefficients along these cuts are also derived and presented in Table~\ref{tab:correlation_filament}.
Notably, across all AR1 and cuts, the highest correlation consistently occurs between 304~\AA\ and Ly$\alpha$, with values exceeding 0.76 and reaching as high as 0.88. This further confirms the physical explanation of the close formation heights between the two wavelengths.

Similarly, we derived the normalized flux along the three cuts, with the results displayed in Figure~\ref{fig:cut_filament}.
From the figure, it is evident that the intensity distributions in the AIA 1600~\AA\ and 1700~\AA\ are highly similar. 
However, these distributions differ significantly from the \lya distribution along Cut2, which is consistent with the relatively lower correlation coefficients presented in Table~\ref{tab:correlation_filament}.
In the case of Cut3, which crosses a bright structure located around 300--350~arcsec along the y-axis, the correlation coefficients are all relatively high.
This is reflected by the presence of prominent, simultaneous peak structures in the intensity distributions across all five channels.
In contrast, at Cut2, a dark footpoint located around 300--320~arcsec may reduce the correlation as flux differences outside the footpoint have a more significant impact. 
For instance, the peak structure near 350~arcsec and the intensity distribution beyond 400~arcsec differ significantly between the 1600~\AA\ and Ly$\alpha$, leading to a low correlation value of 0.172.
Additionally, the correlation between \ha and 304~\AA\ can vary substantially across different cuts, which can be attributed to localized atmospheric structuring and dynamic effects.

\begin{table}
    \caption{The summarized correlation results between different passbands for the four regions (PR, AR1, AR2, QR).}
    \label{tab:all_regions}
    \begin{tabular}{c|c|c|c|c|c}\hline
        Waves & 304 VS \lya & \ha VS \lya & 304 VS \ha & 1600 VS \lya & 1700 VS \lya \\\hline
        PR & 0.530 & 0.317 & 0.333 & $\backslash$ & $\backslash$ \\
        AR1 & 0.845 & 0.222 & 0.130 & 0.414 & 0.281 \\
        AR2 & 0.863 & 0.416 & 0.477 & 0.532 & 0.458 \\
        QR & 0.561 & -0.116 & -0.180 & 0.160 & 0.104 \\
        \hline
    \end{tabular}
\end{table}

Another active region (AR2) shown in Figure~\ref{fig:all-image} is chosen to carry out more comparison. 
This region contains two filaments: one in the upper-right corner, the other in the lower-left region. 
We calculated the correlation coefficients over the AR2, with the results shown in the third row of Table~\ref{tab:all_regions}.
As in AR1, the highest correlation occurs between 304~\AA\ and Ly$\alpha$, again.
Several other correlation results exhibit similar levels of range between 0.41 and 0.54, though all exceed the corresponding values observed in AR1. 
It appears that the more dynamic the active region is, the stronger the correlations are, indicating the influence of dynamic activity on the structural similarity in these passbands.

\subsection{On Quiet Region}

To ensure comprehensive representation, a quasi-simultaneous quiet region (QR) is selected and shown in Figure~\ref{fig:all-image}. The atmosphere structures at 1600~\AA, 1700~\AA, and \ha appear relatively uniform and smooth due to the proximity of the imaging location to the photosphere. However, the 304~\AA\ and \lya images display more inhomogeneous structures. Likewise, the derived correlation coefficients for QR, presented in the fourth row of Table~\ref{tab:all_regions}, are consistent with these observations. Notably, only the correlation between 304~\AA\ and \lya exceeds 0.5, while the others are all below 0.2. 
The correlation between \ha and 304~\AA, as well as between \ha and Ly$\alpha$, is even negative, indicating no significant relationship here. The correlation in the QR is noticeably lower compared to the previous two active regions (AR1, AR2), which aligns with our earlier opinion that the degree of correlation is strongly influenced by the level of activity in the solar structures. More dynamic, active regions tend to exhibit higher correlation values.

\section{Conclusions and Discussions}
\label{sec:discussions and conclusions}

We presented the SRF of the LST/SDI and fitted it with a Gaussian profile, which revealed that the FWHM of the SRF is approximately 85~\AA. As a result, this broad response include contributions from other emission lines, where the \si line is the most prominent. Additionally, we aligned the \lya images captured by SDI and EUI, and compared the image features. 
The calculated correlation coefficient between the two image intensity in the quiet region was 0.86, indicating a strong consistency. 
The polynomial fitting curves showed that the cubic polynomial fit provides the most significant correlation. However, the linear relationship is still the main trend between the intensity of the two images due to the nonlinear terms are all a few orders of magnitude smaller than the linear ones. 

Concerning flares, the \si emission is prominent within the broad SDI \lya waveband. 
We analyzed full-disk spectral profiles ranging from 1203 to 1227~\AA\ observed by SOLSTICE during six M-class and X-class flares and observed that the \si line, as well as the \lya wings, reached a maximum intensity around the SXR peak time of flares. 
The integrated intensity ratio $I$(Si~\textsc{iii})/$I$(Ly$\alpha$) ranged from 1.7\% to 14.6\%, depending on the flare level, from M1.0 to X17.2.
Further analysis revealed a linear empirical relationship between the $I$(Si~\textsc{iii})/$I$(Ly$\alpha$) and SXR flux, described by the function $y=55.6x+0.0175$, derived from flares larger than C-class during 2003 to 2012. 
This near-proportional relationship indicates that the more the SXR flux is, the larger the ratio $I$(Si~\textsc{iii})/$I$(Ly$\alpha$) is. Thus, the \si contribution can not be ignored for large-class flares. 
When accounting for the influence of SRF in SDI observations during flares, more accurate \lya radiation measurements can be obtained by applying the weighting factor $\delta=0.914$ to $I$(Si~\textsc{iii})/$I$(Ly$\alpha$) within the established linear relationship. For M-class flares, this correction induces differences of less than 0.2\%, which remains observationally negligible.

In prominences and quiet regions, the $I$(Si~\textsc{iii})/$I$(Ly$\alpha$) is comparatively smaller, as confirmed by SUMER spectral observations. This finding is consistent with previous conclusions. We obtained the distribution of the integrated intensity ratio using the same SUMER data employed by \cite{curdtSUMERLyLine2010} and \cite{gunarQuietSunHydrogenLymanalpha2020}, and the results are summarized in Table.~\ref{tab:Ratio_SUMER}. None of the ratios exceeds 2\%, and the peak values across different regions follow a similar trend: this ratio increases along with the distance to the disk center within the solar disk. Additionally, the peak values within prominences are significantly lower. To further investigate, we analyzed the center-to-limb variation of the \lya line using full-disk observations from LST/SDI. Consistent with earlier studies \citep{curdtLyMathsfAlpha2008,tianHYDROGENLyaLyv2009}, we found no significant variation in \lya intensity from the disk center to the limb. In contrast, the \si line shows a slight increase in intensity as the distance to the center increases \citep{tianHYDROGENLyaLyv2009}, leading to a rise in the ratio as one approaches the limb. Beyond the limb, however, the ratio decreases due to the more rapid decline in \si radiance compared to Ly$\alpha$. The relative fluxes would be minimally impacted by the discrepancies between observations from different instruments \citep{Greatorex2024SoPh..299..162G}.

Additionally, we compared the \lya image intensity from LST/SDI observations with UV images (304~\AA, 1600~\AA, 1700~\AA) from SDO/AIA and \ha images from CHASE/HIS, and calculated the correlation coefficients between the image intensity across four different regions: PR, AR1, AR2, and QR. The correlation results are presented in Table~\ref{tab:all_regions}.
The highest correlation coefficients are consistently between 304~\AA\ and \lya in all regions. This is expected, as the He~\textsc{ii} 304 \AA\ channel primarily observes the chromosphere and transition region \citep{lemenAtmosphericImagingAssembly2012}, which is relatively closer to the formation height of the \lya line \citep{1981ApJS...45..635V}. In contrast, the 1600~\AA\ and 1700~\AA\ channels are sensitive to the upper photosphere and temperature minimum region \citep{lemenAtmosphericImagingAssembly2012}, while the \ha line traces the structure of the lower atmosphere, including the chromosphere and photosphere \citep{liChineseHaSolar2022}. As a result, the closer formation heights of He~\textsc{ii} 304~\AA\ and the H~\textsc{i} \lya contribute to their strong correlation.
Furthermore, the correlation coefficients in the active regions (AR1 and AR2) are notably higher than those in the QR, likely due to the bright, complex structures abundant in the active regions, which enhances the correlation. In the prominence region (PR), however, the correlation coefficients are all not great.


%

%

%

%
\begin{acks}
We are sincerely grateful to Malcolm Druett for discussing the center-to-limb variation of \lya radiation. We also thank the teams of SORCE/SOLSTICE, SOHO, SDO/AIA, CHASE, and SolO/EUI for their open-data policy. The ASO-S mission is supported by the Strategic Priority Research Program on Space Science, Chinese Academy of Sciences. The SORCE is a NASA-sponsored satellite mission, operated by the Laboratory for Atmospheric and Space Physics (LASP), USA. The SOHO mission is a joint ESA/NASA project. The CHASE mission is supported by China National
Space Administration. SDO is a mission of NASA's Living With a Star Program. SolO is a space mission of international
collaboration between ESA and NASA, operated by ESA.
\end{acks}

\begin{authorcontribution}
Y.L. Li wrote the main manuscript, Y.L. Li and Ping Zhang analyzed most of the data, Y.L. Li generated most of the figures. Y.L. Li and P. Zhang contributed equally to the work and are co-first authors. L. Feng conceived the study, guided the data analyses, and revised the manuscript. Z.Y. Tian analyzed the flares data and generated part of the figures. G.L. Shi helped with some code issues. J.C Xue provided an alignment code for the AIA and CHASE images and gave advice on some physical explanations. Y. Li, J. Tian, B.L. Ying and S.T. Li participated in the discussion of results and provided some suggestions for revisions. K.F. Ji provided an alignment code for the AIA and SDI images. W.Q. Gan is PI of the ASO-S. H. Li and L. Feng are PI and Co-PI of the LST, respectively. All authors reviewed the manuscript.
\end{authorcontribution}
\begin{fundinginformation}
 This work is supported by National Key R\&D Program of China 2022YFF0503003 (2022YFF0503000), the Strategic Priority Research Program of the Chinese Academy of Sciences, Grant No. XDB0560000, NSFC (grant Nos. 12233012, 12203102).
\end{fundinginformation}
%
%
%
\begin{ethics}
\begin{conflict}
The authors declare no competing interests.
\end{conflict}
\end{ethics}

%
%
\bibliographystyle{spr-mp-sola}
\bibliography{paper_cite}
%
%
%
%
\end{sloppypar}
\end{CJK}
\end{document}